\documentclass[%
reprint,
longbibliography,
amsmath,amssymb,
aps,
]{revtex4-1}

\usepackage{color}
\usepackage{textcomp}
\usepackage{hyperref}
\usepackage{graphicx}
\usepackage{dcolumn}
\usepackage{bm}

\usepackage{tikz}

\begin{document}
	
	\preprint{APS/123-QED}
	
	\title{Network neuroscience and the connectomics revolution}
	
	\author{Richard F. Betzel$^{1-4}$}
	\email{rbetzel @ indiana.edu}
	
	\affiliation{
		$^1$Department of Psychological and Brain Sciences, $^2$Cognitive Science Program, $^3$Program in Neuroscience, $^4$Network Science Institute, Indiana University, Bloomington, IN 47405
	}

	
	\date{\today}
	\begin{abstract}
		Connectomics and network neuroscience offer quantitative scientific frameworks for modeling and analyzing networks of structurally and functionally interacting neurons, neuronal populations, and macroscopic brain areas. This shift in perspective and emphasis on distributed brain function has provided fundamental insight into the role played by the brain’s network architecture in cognition, disease, development, and aging. In this chapter, we review the core concepts of human connectomics at the macroscale. From the construction of networks using functional and diffusion MRI data, to their subsequent analysis using methods from network neuroscience, this review highlights key findings, commonly-used methodologies, and discusses several emerging frontiers in connectomics.
	\end{abstract}
	
	\maketitle
	\section*{Introduction}

	This book deals with the topics of connectomics and deep brain stimulation. But what is the \emph{connectome} to begin with? How has the concept emerged and what are the current methods and approaches for mapping and studying it? In this chapter, we address these questions.

	To understand the behavior of a complex system, we need to not only understand how the elements of that system behave in isolation, but how those elements interact with one another and the repertoire of patterns that can emerge from their collective interactions \cite{bassett2011understanding, park2013structural, sporns2011human}. The human mind is one such complex system. It helps us sense and perceive our environment, create (and sometimes forget) memories, and even plays a role in fostering creativity. These types of behaviors are underpinned by our nervous system, which is composed of cells, neuronal populations, and brain areas \cite{betzel2017multi}. Yet, the mind does not emerge from these individual neural elements. Rather, complex behavior emerges from the distributed anatomical and functional networks created by interacting neural elements. To understand the mind, we must first understand the structure and function of brain networks \cite{bassett2017network}.
	
	Historically, neuroscience has focused on properties of nervous systems that can be localized to individual neural elements, such as a cortical region's blood-oxygen-level-dependent (BOLD) activity or its curvature and gyrification. Recently, however, the emphasis has shifted toward studying properties of distributed networks \cite{sporns2018new}. This change in perspective is due in large part to the maturation of network science, which has provided a framework for  mathematical representing and analyzing high-dimensional datasets \cite{newman2006structure, barabasi2016network} and has been applied successfully in other disciplines \cite{wasserman1994advances, jeong2001lethality,barabasi2011network, vidal2011interactome,krioukov2012network}.
	
	These two events, along with data-sharing initiatives and access to high-performance computers, helped create a ``connectomics revolution'', in which neuroscience (and especially human neuroimaging using MRI methods) began to explicitly study structure and function of the brain from a network perspective \cite{biswal2010toward, craddock2013imaging}. Today, the procedures for generating brain networks have been largely automated \cite{esteban2019fmriprep}, and network analyses are addressing fundamental questions in neuroscience concerning the brain's organization \cite{bullmore2009complex}, and how it functions in health \cite{moussa2011changes, shirer2012decoding, cole2014intrinsic, cole2016activity} and disease \cite{xia2018linked, li2020neuroimaging, fornito2015connectomics}, with exquisite detail and personalization \cite{laumann2015functional, poldrack2015long, poldrack2017precision, gordon2017precision, gratton2018functional}.
	
	In this chapter, we cover the basic tenets of connectomics and network neuroscience, focusing on large-scale human neuroimaging. We first introduce the concepts of functional and structural connectivity and outline the procedures for reconstructing these types of data from observations. We then discuss how these data can be modeled as a complex network, emphasizing three ways that network models of biological neural networks have revealed key insight into brain organization or function: structure-function relationships, brain network dynamics, and the brain's system-level architecture. Finally, we cover several frontier topics within connectomics, focusing on generative models, network controllability, and edge-centric approaches.
	
	
	\section*{Structural and functional brain networks}
	
	\begin{figure*}[t]
		\centering
		\includegraphics[width=1\textwidth]{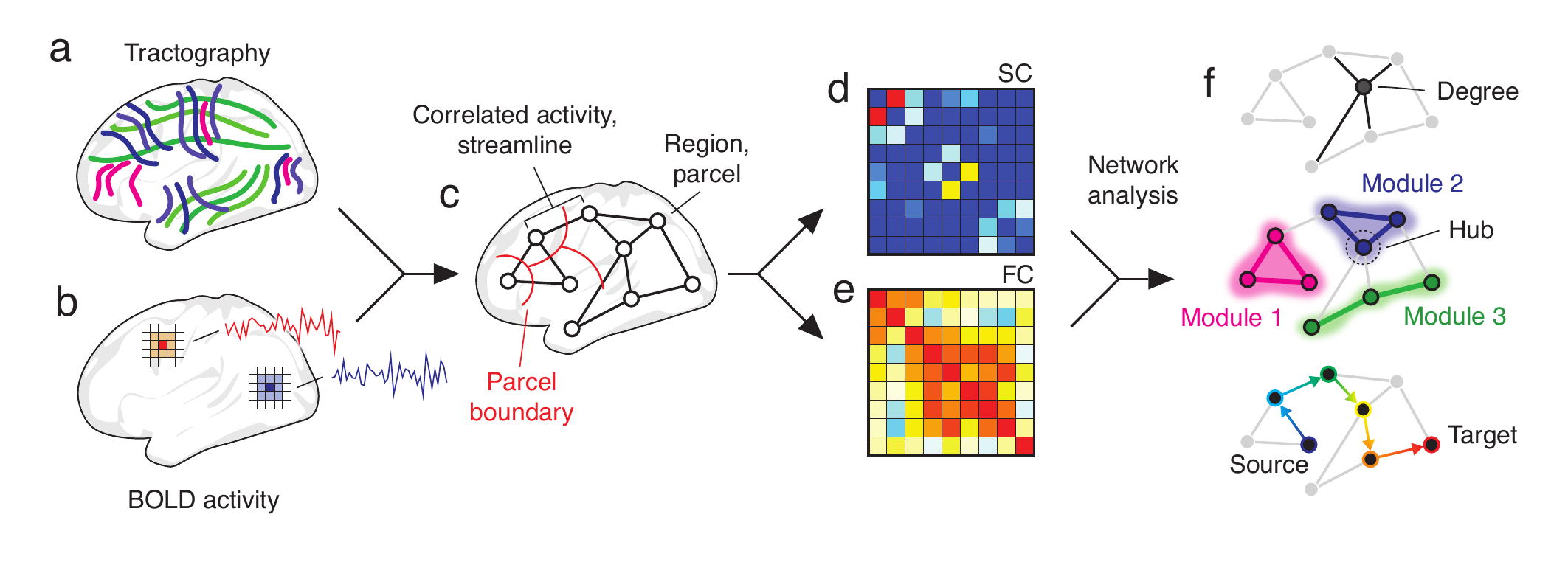}
		\caption{\textbf{Constructing, representing, and analyzing brain networks.} Brain networks are typically constructed from diffusion-weighted and functional MRI data. (\emph{a}) In the case of diffusion MRI, tractography is used to infer white-matter pathways. (\emph{b}) In the case of functional MRI, connectivity is usually defined as the statistical dependence between BOLD time series recorded from voxels or surface vertices. (\emph{c}) The brain is usually divided into parcels or regions so that each parcel is internally homogeneous and composed of voxels/vertices with similar sets of features. Region-by-region structural and functional connectivity matrices (\emph{d} and \emph{e}) are generated by aggregating connections weights according to parcels. In general, structural connectivity (SC) is sparse and its connections are weighted by streamline count, fractional anisotropy, or related measures. Functional connectivity (FC) is fully-weighted and, in the case of Pearson correlations, signed. (\emph{f}) The network representation of SC and FC can be subsequently analyzed, e.g. using network measures at local, meso-, and global scales.} \label{fig1}
	\end{figure*}
	
	Brain networks are maps that tells us which neural elements are connected, which are not, and how strong or efficacious each connection is \cite{sporns2005human}. Although these maps can be created at all spatial scales \cite{schroter2017micro, van2016comparative}, from cells to regions, connectomics has been embraced and proven especially transformative at the macroscale \cite{craddock2013imaging}, where the smallest possible neural elements are voxels, each of which contains roughly $10^6$ neurons (although in practice, voxels get grouped together into larger parcels or regions of interest according to some set of shared features \cite{eickhoff2018imaging, schaefer2018local, fan2016human, desikan2006automated, gordon2016generation, glasser2016multi}). What does it mean for two regions or parcels in the brain to interact with one another \cite{horwitz2003elusive}? In general, the field has converged on three different ``flavors'' or modalities by which two parts of the brain can interact with one another \cite{friston1994functional}.
	
	\subsection*{Flavors of connectivity}
	
	The first measure of interactivity is anatomical or structural connectivity (SC), which describes the physical and material pathways of the brain. At the micro- and meso-scales, SC reflects synaptic coupling between cells or long-distance axonal projections between neuronal populations \cite{oh2014mesoscale, markov2014weighted}. At the macroscale, SC corresponds to large, myelinated white-matter fiber bundles \cite{basser2000vivo}. These bundles can be reconstructed from diffusion weighted MRI data, which measures the apparent direction in which water molecules diffuse in the brain, using a procedure known as ``tractography,'' with many software packages capable of implementing this procedure efficiently (See Chapter 11) \cite{daducci2012connectome, tournier2012mrtrix, yeh2017diffusion, cieslak2020qsiprep} (Fig.~\ref{fig1}a).
	
	Effectively, SC restricts which neural elements can ``talk'' with one another, shaping signaling patterns and the flow of information throughout and inducing statistical dependencies between pairs of neural elements and their recorded activity \cite{honey2009predicting, goni2014resting}. These dependencies are referred to as functional connectivity (FC), in principle, can be quantified using any bivariate measure of statistical similarity \cite{zhou2009matlab, smith2011network, pervaiz2020optimising, horwitz2003elusive, friston1994functional} (Fig.~\ref{fig1}b). These include spectral or wavelet coherence, mutual information, partial correlations, penalized correlations, to name a few. In practice, however, FC between brain regions is almost always measured as the bivariate (Pearson) correlation of the their activity when using functional MRI data (see Chapters 10 and 16).
	
	There are, of course, other ways of measuring the interactivity among neural elements. These include effective connectivity (EC) \cite{friston2011functional, friston2013analysing}, which offers methods like dynamic causal modeling \cite{stephan2010ten}, Granger causality \cite{seth2015granger}, and transfer entropy \cite{schreiber2000measuring}, which measure the directed influence of one neural element on another. In general, estimating true causal relationships between neural elements connectivity is challenging \cite{marinescu2018quasi} and requires experimental manipulation, e.g. observing the effect of a perturbation on the activity of other neural elements, or a specific model of how activity is relayed from one part of the brain to another, usually constrained by SC. Invasive procedures such as deep brain stimulation are especially promising in the context of EC, as they present a unique opportunity to map causal relationships between neural elements \cite{lim2012vivo} and to validate existing causal models.
	
	Other less-common methods for estimating the strength of interactions between neural elements include their structural covariance \cite{mechelli2005structural}, which measures population-level similarity of regions morphological characteristics (although recent advances have made it possible to estimate similar metrics for single subjects \cite{seidlitz2018morphometric}).
	
	\subsection*{Representing brain connectivity}
	
	Irrespective of connection modality, in order to analyze patterns of brain connectivity, we need a way of representing those patterns mathematically. To do this, we turn to graph theory -- a branch of mathematics concerned with the study of networks \cite{rubinov2010complex, bullmore2009complex}. As briefly outlined in Chapter 1, graphs are mathematical representations of a network and consists of two essential ingredients: a set of vertices or nodes that represent the elementary units of a system, and a set of edges or connections that represent interactions between pairs of nodes. In a social system, e.g. an online community such as Twitter, nodes represent users and edges indicate whether one user follows another. In the case of the brain, nodes and edges usually represent brain regions and their structural or functional connections (Fig.~\ref{fig1}c). Although the simplest version of a graph is binary and undirected -- i.e. where edges all have the same weight and encode symmetric relationships -- graphs can also encode information about the weights and asymmetries of connections (those that are and are not reciprocated).
	
	Irrespective of whether a network is weighted/binary, directed/undirected, the pairwise relationships between its nodes can be represented as an adjacency or connectivity matrix. The number of rows and columns in this square matrix is equal to the number of nodes in the network, and its elements represent the presence or absence and weights connections. In an SC matrix, weights usually correspond to streamline counts (the number of reconstructed streamlines between pairs of parcels), their density, or microstructural properties of streamlines, e.g. their fractional anisotropy (Fig.~\ref{fig1}d). Although the precise density of connections in an SC matrix can vary as a function of processing pipelines, structural networks are generally sparse, meaning that of the connections that could exist, relatively few are actually present. Functional connections, as noted earlier, represent bivariate measures of statistical dependence and result in a fully weighted and signed matrix (Fig.~\ref{fig1}e).
	
	Care needs to be taken in preparing connectivity matrices -- the procedures for inferring functional and SC are noisy and prone to both false positives and negatives. It is common, then, as a post-processing step, to impose some threshold on observed connections to retain those with the strongest weights or those least likely to represent false positives \cite{zalesky2016connectome, roberts2017consistency, van2017proportional, betzel2019distance, fallani2017topological}. It is worth noting, however, that thresholding is, in general, not a necessity; in some cases, it may be advantageous to retain all connections. In any case, the decision to threshold or not is non-trivial and will have implications for subsequent analysis of the network \cite{garrison2015stability}. In either case, the thresholded or unthresholded connectivity matrices can then be analyzed further using graph-theoretic network measure to summarize different aspects of a network's organization \cite{rubinov2010complex, bullmore2009complex} (Fig.~\ref{fig1}f). In the next section, we review some of these measures.
	
	\section*{Network measures and analysis}
	
	A network's adjacency matrix provides a succinct description of its topology -- the configuration of connections among nodes. Analyzing brain networks usually involves studying connectivity matrices at different topological scales, ranging from local measurements of individual brain regions and connections to global properties that characterize the entire brain. These measures are usually interpreted in terms of network function. In this section, we describe some of the more common measurements that can be made on a connectivity matrix and provide some interpretations.
	
	\subsection*{Local measures}
	
	Local measures describe properties of individual nodes and edges. For instance, a \emph{node's degree} simply counts the number of incoming and outgoing connections that the node makes (Fig.~\ref{fig1}f). Despite its simplicity, a node's degree is one of the most fundamental measures that can be calculated for a network, influencing other local properties. Nodes with many connections -- putative ``hub'' regions \cite{sporns2007identification, van2013network} -- tend to occupy positions of influence and import within a network (see Chapter 1 for how this may relate to deep brain stimulation).
	
	Centrality measures quantify the importance of a node or an edge with respect to dynamical process taking place on a network or to other structural features of a network \cite{zuo2012network}. The most common of these measures is \emph{betweenness centrality}, which is calculated as the number of shortest paths between pairs of nodes that include a given node. Shortest paths in a network, which are discussed in more detail in the next section, are the shortest sequence of ``hops'' from a source node to a target node in a network. Betweenness centrality can also be calculated for network edges by tallying how many shortest paths involve a given connection. Of course there are other centrality measures that measure the importance of nodes and edges with respect to other processes and structure of a network, e.g. random walks \cite{bonacich1972factoring} and communities \cite{estrada2005subgraph}.
	
	Importantly, most local measures can be calculated for both binary, weighted, directed, and undirected networks \cite{opsahl2010nodevv}. For instance, the weighted analog of node degree is \emph{node strength}, which measures the total weight of all connections incident upon a given node. Similarly, a node's betweenness centrality can be computed based on shortest weighted paths between nodes.
	
	\subsection*{Global measures}
	
	At the opposite extreme are global measures, which quantify properties of the network as a whole. A network's shortest paths structure, for instance, can be used to calculate several common global network measures including \emph{characteristic path length}, which measures the average number of steps in shortest paths \cite{newman2001scientific} (Fig.~\ref{fig1}f). This measure has important functional consequences -- in transportation networks, where goods or information gets delivered from one node to another, shorter paths mean increased efficacy of transmission. In the context of a nervous system, where brain regions need to communicate with one another, shorter paths mean faster transit times and fewer opportunities for a signal to degrade, get transformed, or otherwise attenuate \cite{sporns2004small, achard2006resilient}.
	
	Characteristic path length is an unbounded measure, making it difficult to compare across networks of different sizes. It can also yield a value of $\infty$ if a network is composed of disconnected components (the path length between two disconnected nodes is infinite). To circumvent these issues, one can calculate a the related global measure of \emph{efficiency} \cite{latora2001efficient}. Whereas characteristic path length is the average shortest path between all nodes, efficiency is equal to the average reciprocal shortest path length (1 divided by the length of the shortest path). Consequently, any nodes that are disconnected and have infinite path length now have an efficiency of 0 ($1/\infty$). The efficiency measure is bounded to the interval of $[0,1]$.
	
	Importantly, both path length and efficiency can be extended from binary networks -- where shortest paths correspond to number of hops or steps -- to weighted networks. Some care needs to be taken, however. The algorithms used to map shortest paths seek to minimize a cost function -- the number of steps or total weight along a path. However, connection weights in brain networks usually measure the affinity or strength of connection -- the most efficacious path from a seed to a target is the one that uses strongly-weighted connections. Consequently, shortest weighted paths can be calculated, but require the intermediary step of transforming connection weights from a measure of affinity to length. This is usually accomplished by a negative exponentiation of edge weights, i.e. $L_{ij} = W_{ij}^{-\gamma}$, or by dividing edge weights by their maximum value and taking the negative logarithm, i.e. $L_{ij} = -log_(\frac{W_{ij}}{W_{max}})$. These transformations map weights to length in a monotonic way, so that bigger weights necessarily correspond to smaller lengths. See \cite{van2014neural} for an example of how such global network measures have been used to study effects of deep brain stimulation.
	
	\subsection*{Mesoscale measures}
	
	Situated between the local and global scales is a rich mesoscale \cite{newman2012communities, newman2006modularity, girvan2002community}. Whereas local and global network properties describe features of nodes and edges or the network as a whole, mesoscale measures focus on properties of subnetworks within the larger network, which are thought to support specialized brain function (Fig.~\ref{fig1}f). In the parlance of network science, these subnetworks or clusters are referred to as modules or ``communities,'' in reference to social networks where communities represent groups of individuals. Analysis of mesoscale structure is arguably one of the most active areas of network science \cite{fortunato2016community} (and of connectomics, as well \cite{sporns2016modular}), where it is used for pattern discovery, identification of functional groups, coarse-graining of networks, and for generating additional local and global network statistics.
	
	However, with the exception of trivial examples, a network's mesoscale structure is unknown ahead of time and cannot be determined from visual inspection alone. Instead, mesoscale structure is detected algorithmically using a suite of ``community detection'' algorithms \cite{fortunato2010community}. These algorithms aim to partition a network's nodes or edges into communities, usually so that communities are non-overlapping (although this is not always true \cite{palla2005uncovering, xie2013overlapping, ahn2010link}). The most popular methods in network neuroscience are modularity maximization \cite{newman2004finding} and Infomap \cite{rosvall2008maps}, which seek partitions of nodes that optimize their respective objective functions.
	
	The mesoscale, itself, spans multiple organizational levels, from a division of the network into singleton communities (every node is its own community) to a partition in which the entire network is assigned to the same cluster. Methods like modularity maximization can be parametrically tuned to detect communities of specific sizes between these two limits \cite{reichardt2006statistical} (recent versions of Infomap have a similar ability).
	
	Once a network's community structure has been detected, it can be used to calculate additional local and global network measures. For instance, the objective function, $Q$, used to optimize modularity maximization can be used as an index of how segregated a network's communities are from one another \cite{sporns2016modular, wig2017segregated}. Similarly, we can view communities from the perspective of individual nodes or brain regions, and consider how their connections are distributed within and between modules using measures like \emph{participation coefficient} \cite{guimera2005functional}, which, along with its partner measure, \emph{module degree z-score}, can be used to infer whether a node acts as a hub and whether its influence extends beyond its own module. For instance, nodes with large participation coefficients have connections that span span module boundaries and are positioned to mediate flows between modules \cite{power2011functional}.

	\section*{What have we learned?}
	
	Network neuroscience and connectomics provide a quantitative framework for investigating the brain's organization from a network perspective. Studies employing these methods are beginning to reveal the organization of functional and structural brain networks, shedding light on their operation in health and disease. For instance, like most real-world networks, brains exhibit \emph{small-world} architecture \cite{achard2006resilient, sporns2004small, watts1998collective}, meaning that they have high levels of local clustering to support specialized processing but also topological ``shortcuts'' to reduce the average length of shortest paths, supporting rapid transmission of information across the brain \cite{watts1998collective}. Other studies have demonstrated that brain networks exhibit heterogeneous degree distributions, such that a small number of brain regions make many more connections than others \cite{hagmann2008mapping}. These putative ``hub'' nodes occupy positions of influence and are interlinked to one another, forming a so-called ``rich-club'' of highly-influential brain regions \cite{van2011rich, zamora2010cortical}. Interestingly, rich club nodes are distributed throughout the brain's functional systems, suggesting that these interconnected hubs are positioned to integrate information from different sub-systems and modalities \cite{van2013anatomical}.
	
	While modern connectomics is multifaceted and includes many active areas of research, both applied and theoretical, in the following section we discuss three particular topics that have been especially influential. These include mapping the brain's organization into functionally specialized sub-systems, elucidating the link between structural and FC, and tracking changes in functional network architecture across time.
	
	\subsection*{System-level organization}
	
	Most real-world networks exhibit some form of mesoscale structure, such that nodes and edges can be meaningfully partitioned into sub-networks, modules, and communities \cite{porter2009communities, newman2012communities}. The character of these communities can vary, however. For instance, a network's community structure can form a dense core and sparse periphery \cite{rombach2014core}, or even exhibit disassortative, multi-partite structure where sub-networks are comprised of nodes that are sparsely connected internally, but exhibit many strong connections between groups \cite{murata2009detecting}. The most common type of meso-scale organization, however, is assortative modular structure, where nodes are arranged into sub-networks that are densely connected internally, but weakly connected to one another \cite{newman2004finding, rosvall2008maps}. Indeed, modular structure is ubiquitous in real-world networks \cite{leskovec2009community, girvan2002community} and the development of new methods for detecting and characterizing these types of communities remains an active area of study in network science, broadly \cite{hoffmann2020community, zhang2014scalable, peixoto2014hierarchical}.
	
	Is the brain modular? Why would we expect the brain to exhibit modular organization? Do brains with modules have some advantage that amodular brains are lacking? 
	
	A key feature of modular networks is the near autonomy of their modules, which enables them to develop specialized cognitive and psychological functions \cite{simon1991architecture, espinosa2010specialization, fodor1983modularity}. The autonomy of modules from one another also has implications for network robustness \cite{tran2013relationship}. Suppose a brain were damaged in some way, e.g. as the result of stroke, traumatic brain injury, or due to a deep brain stimulation electrode. In a modular brain, the effect of this damage remains localized to the module in which it originated, impacting only the set of functions supported by that module, but otherwise preserving the broader set of cognitive functions. Additionally, modular structure buffers cryptic variation, rendering brains more evolvable and leading to improved fitness \cite{kirschner1998evolvability}. It helps reduce total cost of wiring and supports a rich repertoire of dynamics \cite{pan2009modularity} and leads to reductions in wiring cost \cite{bassett2010efficient}.
	
	Is there any empirical evidence supporting the hypothesis that brains are modular? An ever-growing number of studies have used clustering and community detection algorithms to uncover the modular structure of real-world structural and functional brain networks \cite{hagmann2008mapping, betzel2013multi, wu2011overlapping, moussa2012consistency, mumford2010detecting, eickhoff2011co, gordon2016generation, meunier2010modular, meunier2009hierarchical, betzel2017modular, bertolero2015modular, akiki2019determining, murphy2016explicitly}. Of these studies, some of the most compelling evidence that the brain exhibits modular structure comes from the analysis of resting-state functional imaging data \cite{power2011functional, yeo2011organization}. In these studies, FC is estimated from fMRI data obtained at rest and clustered -- in the first study, the clustering algorithm Infomap \cite{rosvall2008maps} was applied to thresholded connectivity matrices, and in the second, a modified \emph{k}-means algorithm was applied to FC that was fully-weighted and signed. As expected, these algorithms partitioned parts of the brain into modules. Clustering algorithms, however, can mislead in that they can detect clusters even in networks with no true cluster structure \cite{guimera2004modularity, peel2017ground, hric2014community}. To mitigate this issue, both studies cross-validated their clusters by comparing clusters' the spatial extents with task activation profiles from previous studies. Surprisingly, the boundaries of modules at rest circumscribed the activation patterns, suggesting that the brain's modular structure subtends the same systems that support active cognitive processing.
	
	In another elegant study, Crossley \emph{et al.} \cite{crossley2013cognitive} directly compared the community structure of resting state connectivity with task co-activation patterns. Using data from BrainMap \cite{laird2005brainmap, fox2002mapping}, a database that curates and reports activation locations from many studies, Crossley constructed a co-activation matrix where nodes were connected to one another if they were active under similar conditions. The authors then applied a community detection method to these two matrices, and found a close correspondence between the resulting modules. These observations further suggest that the brain's intrinsic modular structure delineates its functional systems.
	
	While there is an agreed-upon correspondence between communities and the brain's intrinsic functional architecture, there remains many open questions. For instance, what are the ``true'' number of communities? Is there a one-to-one correspondence of communities to functional domains? Do all brain regions form modules? Are all modules segregated and assortative?
	
	Addressing these questions remains an area of active research and one that has required, in some cases, methodological innovation. Take for instance the question concerning the ``true'' number of modules and their respective sizes. Many studies have reported that the brain's modular structure is hierarchical, with smaller communities nested inside of larger communities \cite{meunier2009hierarchical, doucet2011brain}. Even empirical studies have reported divisions of the brain into clusters of vastly different sizes, from large task-positive/-negative bipartitions \cite{golland2008data} to many fine-scaled sub-divisions of these systems \cite{gordon2020default}. However, methods like Infomap and modularity maximization return a partition of the network into modules at a single scale and into a fixed set of communities. How do we reconcile this reality with the expectation that modules should appear hierarchical? One strategy is to employ multi-scale community detection methods. Infomap and modularity maximization both include free parameters that can be tuned to detect different sizes and numbers of communities \cite{betzel2017multi}, effectively resulting in a description of a network's community structure that prioritizes no single scale. Rather, it embraces the brain's inherent multi-scale and hierarchical organization \cite{ashourvan2019multi}.
	
	Community-level analyses of both functional and structural brain networks represent an active area of research and methods development. New techniques like multi-layer modularity maximization \cite{mucha2010community} make it possible to simultaneously cluster data networks from multiple individuals \cite{betzel2019community}, and imaging modalities, and at different points in time \cite{bassett2011dynamic}. Other methods even challenge the basic assumption that the brain' community structure is modular. Stochastic blockmodels make fewer assumptions about the character of communities (both Infomap and modularity maximization can detect only assortative communities), and if core-periphery or disassortative communities exist, are capable of detecting them. Several recent studies using blockmodels have detected non-assortative communities \cite{pavlovic2014stochastic, moyer2015blockmodels, faskowitz2018weighted, faskowitz2020mapping, pavlovic2020multi, gu2020unifying}, showing that by incorporating these features into models, strengthen structure-function relationships and align more closely with patterns of gene co-expression than communities estimated using modularity maximization \cite{betzel2018diversity}.
	
	Meso-scale analyses are also being used to characterize individual differences in cognition \cite{stevens2012functional, baniqued2018brain, bertolero2018mechanistic}, clinical status \cite{alexander2010disrupted, he2018reconfiguration}, and to track developmental \cite{gu2015emergence, baum2017modular} and lifespan changes in brain network structure \cite{chan2014decreased, betzel2014changes, geerligs2015brain}. For instance, a growing number of studies have linked the quality of modules -- quantified using the modularity function -- with behavioral, demographic, and cognitive variables, such as biological age, performance on cognitively demanding tasks. Others have used modularity as prognostic measure, for instance to predict the outcomes therapy and targeted interventions \cite{arnemann2015functional, gallen2016modular, gallen2019brain}.
	
	Others, still, have investigated how the brain's modular structure reconfigures when the brain is engaged in cognitively demanding tasks \cite{cole2014intrinsic}. Converging evidence suggests that a key organizing principle of task-evoked modular structure is that, relative to rest, modules exhibit increased integration and reduced segregation \cite{cohen2016segregation, godwin2015breakdown, gonzalez2018task, betzel2020temporal}. That is, cognitively demanding tasks seem to require that brain modules coordinate with one another, forming strong cross-module links. Methods like stochastic blockmodels report consistent findings, with assortative and segregated modules at rest getting replaced by non-assortative structures like core-periphery structure that, in at least some cases, is task-specific \cite{betzel2018non}.
	
	In summary, investigating the brain's system-level organization from a network science and connectomics perspective has informed multiple domains of neuroscience, from theory to application. This topic remains a productive area of research, and has spurred along methods development and is being applied fruitfully to brain data at all spatial scales \cite{betzel2020organizing, mann2017whole, vanni2017mesoscale}.
	
	\subsection*{Structural basis of activity and connectivity}
	
	\begin{figure*}[t]
		\centering
		\includegraphics[width=1\textwidth]{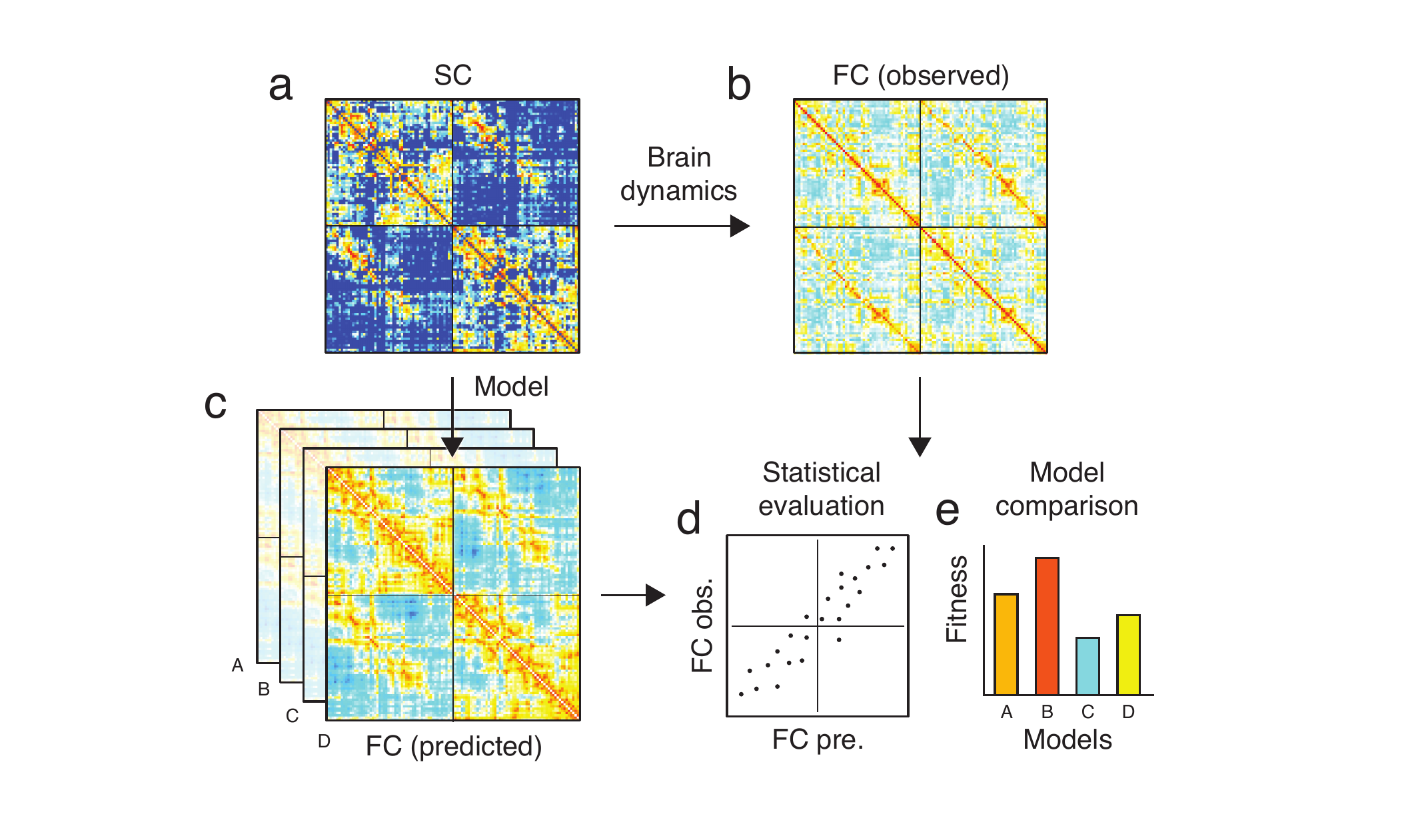}
		\caption{\textbf{Structurally constrained models of FC} Structural connections (\emph{a}) constrain brain dynamics, which induces correlated brain activity that is measured as FC (\emph{b}). To better understand this structure-function relationship, one build models of brain dynamics that yield, as output, synthetic patterns of FC. (\emph{c}) This framework makes it possible to test different models corresponding to distinct hypotheses about brain dynamics/communication with varying levels of complexity and biophysical plausibility. (\emph{d}) Model performance can be quantified as a correlation between observed and predicted FC. (\emph{e}) Models can be compared to one another using their performance as a measure of fitness to identify the optimal dynamics.} \label{fig2}
	\end{figure*}
	
	One of the central questions in biology concerns the relationship of an organism structure with the kinds of functions it can perform. In the context of network neuroscience we can ask a similar question -- how does the brain's anatomical structure shape patterns of functional coupling between distant sites in the brain \cite{suarez2020linking}? Intuitively, we think of anatomical connectivity as a set of constraints -- some parts of the brain are directly coupled to one another via white matter or axonal projections, others are not. Those that are connected can readily exchange information with or signal to one another -- those that lack direct connections can only communicate indirectly, through multi-step pathways, thereby shaping the correlation structure of brain activity \cite{avena2018communication}. This coupling of function to the underlying anatomy is a fundamental feature of brain networks that, when disrupted or perturbed, can give rise to maladaptive behavior and disease \cite{fornito2015reconciling}. A disruption of structure-function coupling may be exactly what we should aim at retuning with deep brain stimulation (see Chapter 1).
	
	Indeed, there is considerable evidence -- both empirical and \emph{in silico} demonstrating that brain SC is causally related to patterns of functional coupling. Some of the strongest evidence comes from lesion studies, where direct structural insults lead to acute changes in FC. One of the clearest examples is a study by O'Reilly et al \cite{o2013causal}, in which resting state FC was estimated for macaque monkeys before and after surgically removal the collosal fibers between the left and right hemispheres. This had the acute effect of reducing the magnitude of interhemispheric functional connections, presumably because those regions can not longer exchange information or synchronize their activity with one another.
	
	Other evidence that SC shapes functional coupling comes from \emph{in silico} studies, where SC is considered part of a networked dynamical system (Fig.~\ref{fig2}a; also see Chapter 26 for an outlook of how these concepts are applied to the field of deep brain stimulation). The elements of a dynamical system -- nodes in the network -- have internal states that represent their level of activity, e.g. membrane potential or BOLD amplitude, and are free to evolve over time according to some internal dynamics. When these elements are coupled to one another -- i.e. as parts of a network -- then the evolution of each element's state depends on the states of its neighbors. Consequently, the constraints imposed by SC induce correlations in the change of nodes' states across time, leading to observed patterns of FC (Fig.~\ref{fig2}b).

	These kinds of models often include biophysical parameters, e.g. ``neural mass models'' include populations of excitatory and inhibitory neurons, channel conductances, temporally lagged propagation of signals, etc.. Although the inclusion of these parameters helps preserves some neurobiological plausibility, it comes with a cost. First, solving the system of differential equations is computationally expensive, and can entail hours of run-time on high-performance computers to generate even a few minutes of data (although new software packages promise to reduce the runtimes \cite{heitmann2018brain}). Second, the correlation structure of the synthetic time series is only modestly similar to its empirical counterpart, suggesting that the biophysical models may be missing key components or parameters \cite{honey2009predicting}. This high computational cost for relatively low variance explained has prompted the exploration of simpler and more easily interpretable models based on analytically tractable, stylized dynamics that represent putative policies for interregional communication between brain regions \cite{avena2018communication}.
	
	Unlike the biophysical models, where the output is synthetic time series, communication models generate for all pairs of regions a measure of communication capacity or the ease with which two brain regions can communicate with one another, resulting in a square node-by-node matrix. For instance, suppose we hypothesize that brain regions ``communicate'' by routing signals through the network's shortest path structure. As a coarse test of this hypothesis, we could compute the length of shortest paths between all pairs of brain regions and compare the elements of the resulting shortest paths matrix with those of the empirical FC matrix. In general, stronger correspondence between the two matrices can be interpreted as evidence supporting the hypothesis that shortest paths are important for brain communication.
	
	To take advantage of shortest paths for signaling, the ``signals'' would need to have global knowledge of the entire network and its shortest path structure. At the opposite extreme are decentralized processes like diffusion models in which a particle or random walker (representing some quanta of information) moves over the network randomly, hopping unbiased from one node to another \cite{abdelnour2014network, abdelnour2018functional, goni2013exploring}. From this process, one can compute measures like the mean first passage time that a particle starting at one node is expected to reach another, and compare these measures against FC. Other simple models include those based on communicability in which nodes communicate along all paths but discount longer paths exponentially \cite{crofts2009weighted}, oscillator models where nodes update their phase based on those of their neighbors \cite{schmidt2015kuramoto}, and search information where paths between nodes are quantified based on their hiddenness (the amount of information necessary for a random walker to follow the shortest path without error) \cite{goni2014resting}.
	
	These simple models have allowed researchers to explore and compare a veritable zoo of possible communication policies, from multi-path communication \cite{avena2017path} to models of epidemic spreading that have been repurposed for brain imaging data \cite{meier2017epidemic, raj2018models, mivsic2015cooperative} to decentralized and greedy navigation models \cite{seguin2019inferring, seguin2018navigation, allard2020navigable}, to information-based models \cite{amico2019towards, avena2019spectrum} (Fig.~\ref{fig2}c). Because they are based on analytically tractable measures, these models carry a low computational cost and can be easily implemented and their parameters optimized. Critically, these models tend to outperform the more neurobiologically plausible models in terms of how well they predict empirical FC \cite{goni2014resting} (Fig.~\ref{fig2}d,e).
	
	Collectively, there is considerable empirical and \emph{in silico} evidence suggesting that SC plays a causal role in shaping patterns of functional coupling. This evidence has prompted many studies to consider situations where normative structure-function coupling is perturbed. For example, during task performance \cite{hermundstad2013structural}, in neuropsychiatric disorders \cite{cocchi2014disruption}, or in human development \cite{baum2020development, betzel2014changes}. Indeed, variation in structure-function coupling has been associated with individual differences to performance on tasks, subject demographics, and traits \cite{seguin2020network}.
	
	There remain many open questions and challenges, however. For instance, the measured association between SC and FC likely depends on other factors, including the spatial embedding of the brain -- the nearer two regions are to one another, the more likely they are to be connected by both modalities \cite{betzel2019structural, esfahlani2020space}. This mutual dependence on distance makes assessing the true strength of structure-function coupling difficult. Other challenges are more fundamental. The veracity of SC and FC matrices remains a source of concern in most analyses. SC, especially, suffers from known biases that prevent accurate inference of certain white-matter tracts \cite{maier2017challenge, reveley2015superficial}. FC, too, exhibits a peculiar set of biases, including variation attributable to head motion in the scanner \cite{power2012spurious}. More fundamentally, the \emph{de facto} measure of functional coupling -- the Pearson correlation -- represents one of many possible measures of connectivity, and the strength of structure-function relationships will, in general, vary as a function of these measures.

	\subsection*{Connectome dynamics}
	
	\begin{figure*}[t]
		\centering
		\includegraphics[width=1\textwidth]{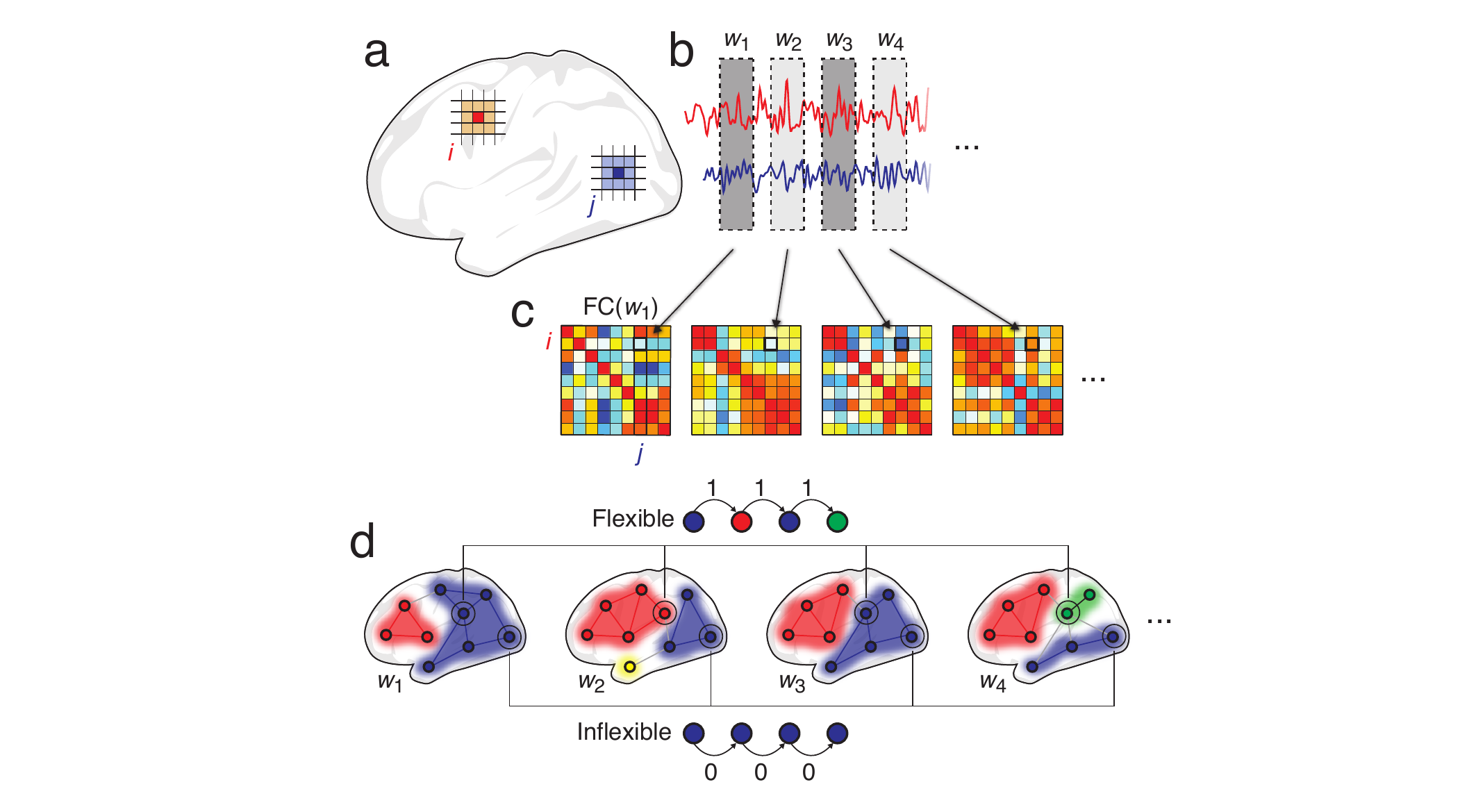}
		\caption{\textbf{Time-varying FC and flexibility.} To study changes in network organization across time, sliding window methods are routinely used to obtain time-varying estimates of FC. (\emph{a}) This involves extracting time courses from different parts of the brain. (\emph{b}) Time series are then ``windowed'' and FC weights are estimated using only the points that fall within a given window. For a pair of voxels/vertices/parcels, this results in a sequence of FC estimates that are temporally localized to that window. (\emph{c}) This procedure can be repeated for all pairs of brain regions, resulting in a time series of FC matrices. (\emph{d}) Although time-varying FC can be analyzed using many different methods, one common approach is based on reconfiguration of network modules to estimate network flexibility. This procedure involves constructing a ``multi-layer network'' from the time-varying matrices and clustering all time points simultaneously. Flexibility is calculated as the frequency with which a node changes its modular affiliation from one time point to the next. Here, we show examples of a node that is flexible (\emph{top}) and another that is inflexible (\emph{bottom}).} \label{fig3}
	\end{figure*}
	
	Among the profound realizations of the connectomics revolution is that brains are never quiet, even during cognitive rest \cite{deco2011emerging}. Rather, the human brain is in constant transit, traversing a high-dimensional landscape of activity and connectivity patterns over time. Time-varying changes in FC, in particular, have attracted considerable interest from the network neuroscience community \cite{hutchison2013dynamic}, where it is generally accepted transient coupling and uncoupling of neural elements across time reflects changes in cognitive state and that by studying these changes, we can glean new insight into principles by which functional networks support psychological processes on fast timescales.
	
	Many methods have been proposed to studying dynamic or time-varying FC \cite{monti2014estimating}, including frequency-based decompositions \cite{chang2010time}, instantaneous co-fluctuation analysis \cite{liu2013time}, and model-based approaches \cite{lindquist2014evaluating}. The most common, however, is the sliding-window approach \cite{hindriks2016can}. This entails specifying a ``window'' of some length comprising temporally contiguous time points (Fig.~\ref{fig3}a,b). Then, temporally resolved FC is calculated using only those time points within a given window and the window advanced some fixed number of time points where the procedure gets repeated. The result is a sequence of networks corresponding to FC patterns at different points in time (Fig.~\ref{fig3}c). These networks can then be analyzed to gain insight into network dynamics -- how the topological organization of the brain fluctuates across time.
	
	Although time-varying FC can be analyzed using many different approaches to quantify the stability and variability of patterns across time, one popular line of inquiry involves assigning time points to clusters based on the similarity FC patterns  \cite{kringelbach2020brain, calhoun2014chronnectome}. This procedure yields a set of brain states -- cluster centroids -- that describe low-dimensional space that the brain traverses over time. These states can be summarized using a rich set of statistics that include the relative frequencies of each state and the probabilities of transitioning from one state to another. This approach has been used extensively and has helped link time-varying connectivity to arousal and wakefulness \cite{young2017dynamic, allen2014tracking}, attention \cite{fong2019dynamic}, and state of consciousness \cite{hutchison2013resting, kucyi2014dynamic, barttfeld2015signature}, and has been applied in a clinical context to generate biomarkers of disease and neuropsychiatric disorders \cite{damaraju2014dynamic}.
	
	Another powerful approach for studying time-varying connectivity is to track changes in communities across time using multi-layer network models \cite{kivela2014multilayer} (Fig.~\ref{fig3}d). In this approach, FC patterns at different time points are treated as ``layers'' in a single multi-layer network, where each layer is weakly coupled to its temporally adjacent neighboring layers \cite{mucha2010community}. When the multi-layer network is clustered, all layers are clustered simultaneously and their cluster labels preserved. This makes it possible to directly compare nodes' cluster labels across temporally adjacent layers to identify flexible nodes that switch their cluster assignments frequently and those that maintain stable assignments across time \cite{bassett2013robust}. This approach can be used to track the formation, persistence, and dissolution of communities across time \cite{mattar2015functional, bassett2011dynamic}. Flexibility can calculated  both at the whole-brain and regional levels and in previous studies has proven to be a potent biomarker of cognition and disease. A growing number of studies have linked patterns of flexibility with learning \cite{bassett2011dynamic, gerraty2018dynamic, telesford2016detection}, executive function \cite{braun2015dynamic}, mood and affective state \cite{betzel2017positive}, and working memory \cite{finc2020dynamic}. Other studies have linked flexibility to psychiatric disorders and schizophrenia \cite{braun2016dynamic}.
	
	Although analyses of time-varying FC using the sliding window approach is common, this procedure requires the user to make several non-trivial processing decisions along. These include deciding upon window length and whether it should be tapered or not \cite{zalesky2014time} and how far the window should be advanced at each step \cite{zhang2016choosing}. Although varying the length of is a simple way to study network changes at different timescales, window length also has an impact on sampling variability. This leads to a tradeoff between timescale and errors, where shorter windows can reveal faster time-varying changes in FC but with a greater proportion of false positives and negatives \cite{zalesky2015towards, leonardi2015spurious}.
	
	Other issues challenge the very notion that time-varying FC could be measured using fMRI methods. These studies argue that, due to the slow and indirect measures obtained using fMRI, the variability observed in time-varying FC analyses is consistent with the sampling variability from a temporally stationary and unchanging correlation structure \cite{laumann2017stability, liegeois2017interpreting}. This observation necessitates that any estimate of time-varying connectivity be compared against a stationary null model.
	
	In summary, time-varying network analyses are being applied widely to study changes in network structure over shorter timescales. The results of these analyses are providing insight into the principles by which brain networks reconfigure across time. Work in this area has proven controversial at times \cite{lurie2020questions}, but continues to strengthen the link between fast changes in network topology and fluctuations in cognitive state. 
	
	\section*{What does the future hold?}
	
	Like any young field, the full potential of connectomics and network neuroscience remains untapped, with exciting new work taking place on many frontiers. In this section, we review several of these areas, focusing on topics of generative modeling, network control, and edge-centric approaches.
	
	\subsection*{Generative modeling}
	
	\begin{figure*}[t]
		\centering
		\includegraphics[width=1\textwidth]{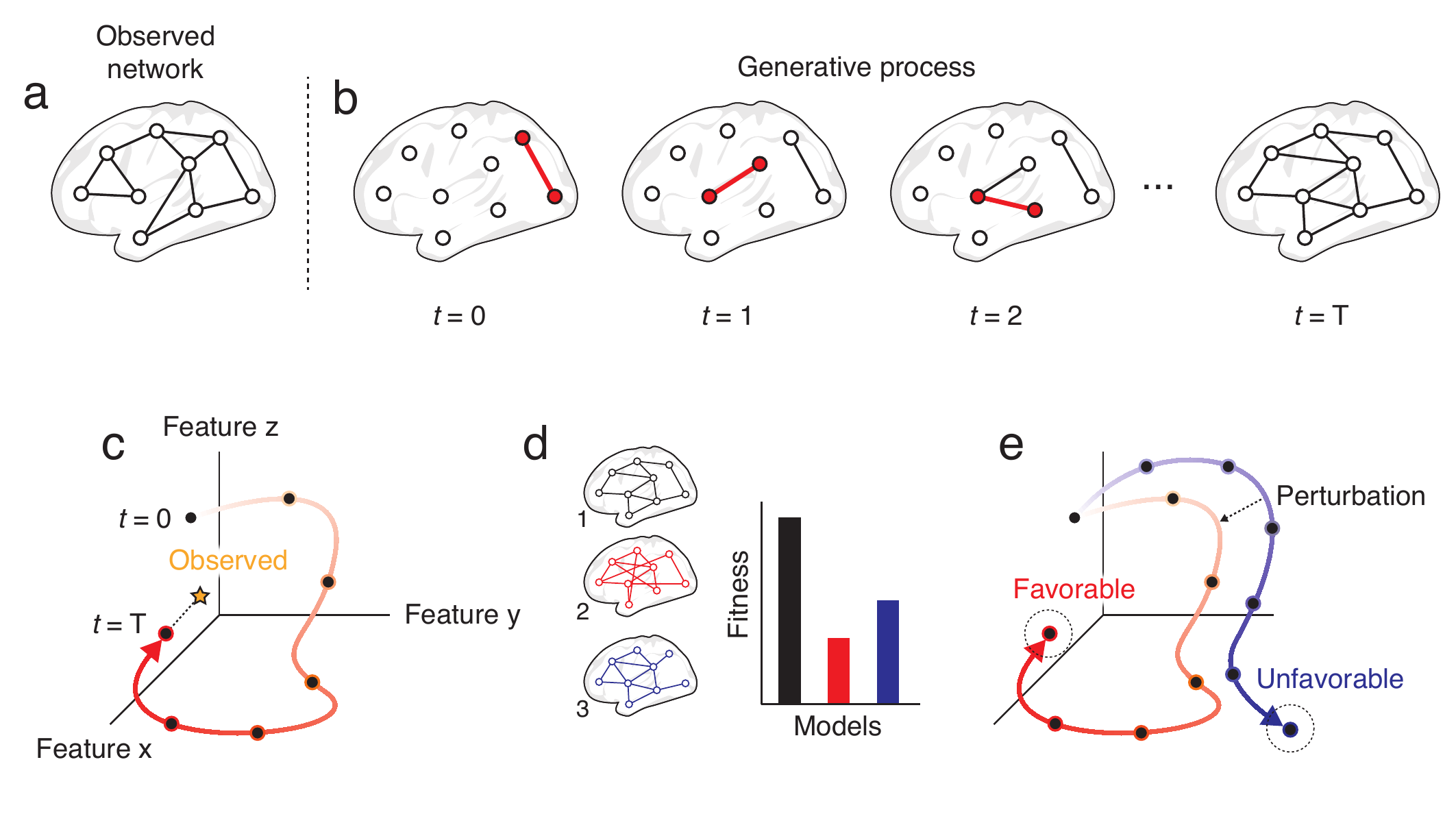}
		\caption{\textbf{Generative models of connectomes.} Network generative models aim to uncover the wiring rules that give rise to an observed brain network (\emph{a}). (\emph{b}) This is accomplished by proposing and testing different models -- generative processes -- and evaluating how well they match features of the observed network. The generative processes can be stylized and take place on non-biological timescales or they can incorporate detailed neurophysiological details and take place over developmental time. (\emph{c}) This process traces a trajectory through a state space, where the outcome -- a synthetic brain network -- is compared to the observed network using some measure of distance/similarity. The generative modeling framework allows for \emph{in silico} explorations of interventions. For instance, if two networks evolve on similar trajectories -- one to a favorable, healthy state and the other to an unfavorable, disease state -- one can use explore targeted perturbations of the unfavorable trajectory to identify strategies that would ``drive'' the brain back towards the healthy state. (\emph{e}) Different generative models that test distinct hypotheses about network growth and evolution can be instantiated and their fitness -- how well they match features of the observed network -- can be compared to gain insight into the wiring rules and principles that shape brain network organization.} \label{fig4}
	\end{figure*}
	
	The appeal of network neuroscience stems in part from the ability to quantify different aspects of brain organization using an ever-growing suite of summary statistics. As noted earlier, these measures can diagnose influential nodes, discover communities, and assess a network's capacity for efficient information transmission. However, network summary statistics are just that --  they \emph{describe} and summarize a network. That is, summary statistics offer no explanation for \emph{why} a network has a certain feature or \emph{how} that feature came to exist in the first place. To understand the processes, rules, and algorithms by which networks grow and evolve requires building, testing, and validating generative models \cite{betzel2017generative}.
	
	Generative modeling has a rich history in network science with the foundational Watts-Strogatz \cite{watts1998collective} and Barabasi-Albert models \cite{barabasi1999emergence} models serving as the clearest examples. While these models posit mechanisms by which small-worldness and hubs emerge, in general, generative models work by identifying a desired set of properties for a network to have, and work backwards to identify the processes that yields networks with those properties (Fig.~\ref{fig4}a,b,c,d).
	
	As with other modeling exercises, parsimony plays an important role, and generative modeling studies usually aim to discover ``simple'' wiring rules that can accurately replicate properties of real brain networks. These rules, then, can be interpreted as  drivers of the brain's network organization. Any additional features that emerge incidentally and as a consequence of these rules can be thought of as ``spandrels'' and useful byproducts of the overarching generative mechanism \cite{rubinov2016constraints}.
	
	The space of possible generative models for brain networks is massive. One way that we can impose some order on this space to categorize models according to how they deal with time. On one extreme are models in which the network grows and evolves over a biologically meaningful timescale, with the aim of recapitulating a specific developmental trajectory. These kinds of models need to be carefully calibrated against observations and include other neurobiological details, but can then be used to assess how perturbations alter developmental trajectories (leading to maladaptive brain networks) and to explore strategies for driving a brain back onto a healthy trajectory, all \emph{in silico} (Fig.~\ref{fig4}e). A good example of this type of model is the work by Nicosia et al \cite{nicosia2013phase} that modeled the growth of the nematode, \emph{C. elegans}. \emph{C. elegans} is unique in that we have detailed information about the birth times of its roughly 300 neurons, which can then be incorporated into a generative growth model with the aim of reproducing the network of the adult \emph{C. elegans}. The authors demonstrated that a model based on birth times combined and wiring rules that penalize the formation of long connections and reward nodes with high degree could account for many of the properties of the adult \emph{C. elegans} connectome, including bi-phasic growth rate.
	
	At the opposite extreme are models for which time plays no role, like stochastic blockmodels, or deal with time in a non-biological way. The Barabasi-Albert model is a good example, in that the network grows according to a set of simple rules, adding nodes and edges over a series of steps, but where the timescale is arbitrary. While these types of models lack biological realism, they can be used to gain insight into the underlying principles that shape network organization. For instance, several recent network neuroscience studies have investigated quasi-dynamic models, in which edges are gradually added to a set of brain regions according to a probabilistic growth rule \cite{vertes2012simple, betzel2016generative}. In these studies, the authors tested many possible rules based on the spatial layout of the network and its topology, comparing their abilities to account for organizational features of empirical brain networks. In both cases, the authors found that the optimal model -- the one that consistently produced synthetic networks with features similar to real-world brain networks -- penalized the formation of long-distance connections but increased the likelihood of connections forming between nodes with similar connectivity profiles. Although the timescale of these models lacked any correspondence with neurodevelopment, their results were consistent with the generative model of \emph{C. elegans}.
	
	Generative models have been instrumental in uncovering potential organizing principles of brain networks. Among the key findings is the role played by spatial relationships in shaping the brain's network structure \cite{ercsey2013predictive, horvat2016spatial, stiso2018spatial}. Brains must be enclosed in a limited volume and their connections require material to form and energy to maintain and use. Shorter connections, therefore, lead to reductions in volume and wiring cost, and lead to increased fitness. Although some studies maintain that the drive to reduce these costs is sufficient to explain brain network organization in its entirety, most agree that this drive needs to be counter-balanced by some opposing force to account for costly features of brain networks that involve long-distance connections, e.g. the presence of hubs and rich-clubs \cite{betzel2018specificity, arnatkeviciute2020genetic, zhang2019generative}.
	
	The generative modeling approach opens up new, exciting strategies for studying brain networks and for understanding their roles in health and disease. First, generative models help discover the drivers of network organization, which helps refocus attention onto those features and away from ``spandrels'' \cite{rubinov2016constraints}, which may help in the generation of increasingly sensitive and appropriate biomarkers. Second, generative models can be fit to subject-specific data and their parameters used to study individual differences. For instance, in \cite{betzel2016generative} the authors demonstrated the parameters governing spatial constraints varied with age, while \cite{zhang2019generative} used similar models and demonstrated that their parameters were correlated with polygenic risk of schizophrenia and cognitive performance. Lastly, generative models can provide insight into human development by incorporating additional neurobiological and developmental details into their wiring rules \cite{akarca2020generative}.
	
	\subsection*{Network control}
	
	In order to meet ongoing cognitive demands, the human brain must seamlessly transition from one brain state to another in order. How does the brain accomplish this? How are these transitions supported by the underlying anatomical connectivity? These types of questions can be addressed using the \emph{network control framework} \cite{liu2011controllability, cornelius2013realistic, liu2016control}. This approach presupposes that the brain is a networked dynamical system and that, in the absence of any intervention, the activity of each brain region evolves over time according to its own state and the states of connected neighbors, tracing out a trajectory over time. However, this trajectory can be altered by exogenous time-varying input signals that are ``injected'' into brain region. As a result, these signals can drive brain activity, causing it to deviate from its passive trajectory. Network control theory asks the question of whether its possible to tailor these input signals so that instead of simply following a \emph{different} trajectory, the brain is directed along a specific, predefined trajectory using the lowest amplitude control signals possible.
	
	In this way, network control offers an elegant and mathematically tractable framework that naturally links brain connectivity, dynamics, and activity. On one hand, it can be used to yield a set of local and global network measures. These include brain region's control profiles -- their abilities to drive the brain into certain classes of brain states, e.g. those that are hard to reach and require significant amounts of effort to reach \emph{versus} those that are easier to reach, requiring little effort \cite{pasqualetti2014controllability, gu2015controllability, karrer2020practical}. We can also compute global measures that reveal whether a network is or is not controllable -- i.e. whether its topology and dynamics allow it to be driven into any arbitrary state. Like other network measures, control metrics can be calculated for individual subjects and used to study inter-individual or group-level differences \cite{cornblath2019sex, medaglia2018network, beynel2020structural}, and have been used to reveal changes across development \cite{tang2017developmental, lee2020heritability}, differences between control and clinical populations \cite{jeganathan2018fronto}, and to understand that structural and dynamic underpinnings of creative ability \cite{kenett2018driving}.
	
	In reality, most brain states are never visited. That is, they correspond to patterns of activity that represent non-functional ``noise'' or are actively dangerous to visit, e.g. seizure or epileptic activity. To make control measures more informative and realistic, one should focus exclusively on transitions between brain states that are actively visited, and ignore those that, for one reason or another, are not. One way to do this is to leverage the \emph{optimal control framework}. Instead of considering transitions to and from all possible states, the optimal control framework considers transitions between specific pairs of states: the initial state in which the system begins and the target state that it is trying to reach. Given a set of control nodes through which input signals are injected, optimal control delivers the time-varying inputs that drive the brain from the initial state to its target using as little effort as possible. The amount of effort -- the control energy -- depends on the character of the initial and target states, but also the topology of the underlying structural network \cite{kim2018role, gu2017optimal, betzel2016optimally, yan2017network}. The optimal control framework yields a series of statistics that can be compared across individuals or between groups. These include nodal trajectories, input signals, and the control energy needed to support the transition. These measures have been used in several applied contexts. For instance, \cite{cornblath2020temporal} calculated brain states as clusters of brain activity patterns, and demonstrated that the most frequent state transitions required the smallest amounts of energy, suggesting that the brain's white-matter network supports its dynamic trajectory through state space. Similarly, \cite{braun2019brain} computed the energy required to transition to and from activity states during a working memory task, and linked the amount of energy to expression of dopamine receptors.
	
	 On one hand, the very premise of network control -- that we can selectively drive the brain into different patterns of activity with distinct cognitive and behavioral analogs -- seems like science fiction \cite{medaglia2017mind}. On the contrary -- network control theory has real-world applications and practical implications. For instance, with the advent of implantable neuromodulatory devices, the widespread use of non-invasive stimulation (e.g. transcranial magnetic stimulation), and the practice of optogenetic control of population-level activity, new theory is necessary to model and predict the effect of targeted stimulation. Nonetheless, there remain some practical hurdles to using network control. To date, most applications of network control to large-scale brain networks assume that brain activity is described by a linear dynamics. While the assumption helps simplify the mathematics of control \cite{karrer2020practical}, linear \cite{cole2016activity}, controlling non-linear dynamics is considerably more challenging than linear control \cite{whalen2015observability, jiang2019irrelevance}. A second issue concerns the relationship of controllability measures with more easily obtained, traditional network measures, e.g. nodal degree. Several articles have demonstrated that these measures are highly co-linear, suggesting that what determines a node's control profile is not the topology of the network, low-level features of nodes, e.g. the number of connections they make. The implication is that the control profiles of a real and randomized network would be nearly identical \cite{tu2018warnings, suweis2019brain}.
	 
	 Nonetheless, network control represents an exciting new frontier in connectomics. It opens up the tantalizing possibility of selectively driving whole-brain patterns of activity using knowledge of the underlying structural connections. While the framework is mostly relegated to exercises in theory, new methods for manipulating brain activity, which include deep brain stimulation, will provide cases where theory can be carefully tested and further refined.
	
	\subsection*{Edge-centric connectomics}
	
	\begin{figure*}[t]
		\centering
		\includegraphics[width=1\textwidth]{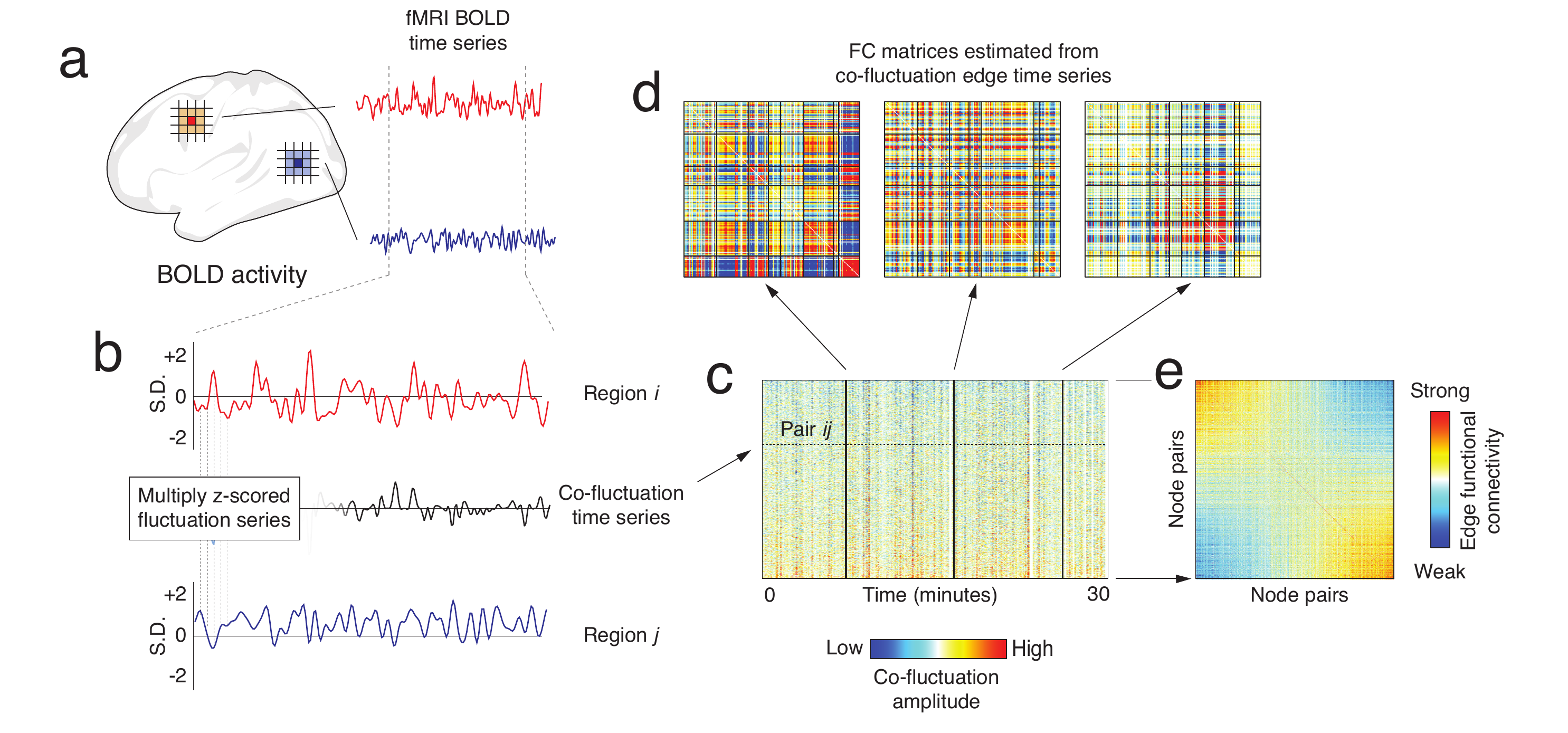}
		\caption{\textbf{Edge-centric modeling and analysis of neuroimaging data.} Edge-centric models shift focus away from individual brain regions and their activity onto pairs of regions and their coactivity. (\emph{a}) This is accomplished by first extracting and z-scoring regional/parcel time series, which are incidentally the first two steps in computing fFC. (\emph{b}) For every pair of time series, we compute the element-wise product of their activity, resulting in a co-activity or co-fluctuation time series. (\emph{c}) This procedure is carried out for all pairs of brain regions, yielding an matrix of ``edge'' time series. (\emph{d}) Each column of the edge time series matrix represents a pattern of co-activity expressed at a specific instant. These patterns can be reshaped into a region-by-region matrix and analyzed as a functional network. (\emph{e}) Edge time series can also be used to create a novel type of connectivity matrix: ``edge functional connectivity'' (eFC). This is accomplished by computing the pairwise similarity between all possible pairs of edge time series. Unlike nodal FC, which results in a region-by-region matrix, the eFC matrix has dimensions of node pairs by node pairs, i.e. edge by edge. This matrix can be analyzed further, e.g. clustered to reveal overlapping communities.} \label{fig5}
	\end{figure*}
	
	Historically, brain networks have been modeled and analyzed using ``node-centric'' networks whose the nodes and edges represent neurons, populations, or areas, and functional or structural connections. This model has been the workhorse of network neuroscience and is at the core of virtually every connectomics discovery to date. Recently, however, several groups have proposed extending the node-centric model to characterize \emph{higher-order} interactions in a network \cite{ahn2010link, evans2009line, owen2019high}. This means shifting focus away from interactions between brain regions and neural elements, and building altogether new classes of networks that focus on the interactions between functional and structural connections.
	
	The two most commonly discussed approaches for generating higher-order ``edge-centric'' networks involve transforming traditional node-by-node connectivity matrices into higher-dimensional edge-by-edge ``line graphs'' \cite{evans2009line} and ``edge similarity'' matrices \cite{ahn2010link} that quantify how strongly pairs of \emph{edges} in the network interact with one another. While these approaches have been applied successfully in other disciplines, their application in neuroscience has been limited. To date, only one study has applied edge-centric methods to empirical brain networks, using line graphs to study the overlapping community structure of SC data \cite{de2014edge}. 
	
	Recently, however, several papers presented a new method for generating higher-order networks, in the process making them easier to apply to neuroimaging data and opening up new opportunities for studying the brain's higher-order organization and dynamics \cite{faskowitz2019edge, esfahlani2020high, jo2020diversity, jo2020subject, sporns2020dynamic}. The new edge-centric model is based on time series and the bivariate Pearson correlation -- the measure routinely used to quantify FC. Whereas the Pearson correlation of two regions' activity is calculated as the mean element-wise product of their standardized (z-scored) time series (Fig.~\ref{fig5}a,b), the edge-centric model omits the averaging step, yielding a co-fluctuation time series for every \emph{pair} of nodes (edge). Because FC is often interpreted as an index of communication between two brain regions, we can think of each edge time series as a time-resolved account of that ``conversation''.
	
	This procedure can be carried out for every pair of brain regions -- all the edges in the network -- to yield a complete set of co-fluctuation time series (also referred to as ``edge time series''), which has useful properties (Fig.~\ref{fig5}c). First, we can view edge time series as a temporal decomposition of FC; the temporal average of the co-fluctuation time series for any pair of nodes is \emph{exactly} the weight of the functional connection between those nodes. With this decomposition, then, we can assess the contribution to average FC of individual frames. Additionally, if we ``slice'' edge time series at a given point in time we obtain the instantaneous co-fluctuation pattern for all pairs of edges, which can be reshaped into a region-by-region matrix and analyzed as a network (Fig.~\ref{fig5}d). Thus, edge time series quantify time-varying changes in network structure without using a sliding window, resulting in improved temporal resolution.
	
	Interestingly, edge time series appear bursty -- i.e. the amplitude is weak at most points in time, but over very short intervals becomes extremely strong \cite{esfahlani2020high}. Even more interesting, the times at which these bursts occur tend to be correlated across edges, resulting time-varying networks where most connections are weak, but are occasionally punctuated by rare and short-lived high-amplitude ``events.'' Although events occur infrequently, they explain a disproportionate amount of variance in static FC, account for its system-level structure, and contain subject-specific information not present in lower-amplitude frames.
	
	Edge time series can also be used to estimate the higher-order construct of \emph{edge functional connectivity} (eFC). eFC is obtained by computing the similarity between pairs of edge time series, yielding an edge-by-edge connectivity matrix analogous to the line graphs and link similarity networks but is fully weighted and signed \cite{ahn2010link, evans2009line} (Fig.~\ref{fig5}e). Like traditional node-centric networks, eFC can be analyzed using tools from network science to identify important or influential edges in the network. The principal advantage of eFC, however, is when community detection algorithms are used to partition edges into communities of cohesively fluctuating edge time series. In this context, edges are assigned to a single community, but because each node is affiliated with many edges, each of which can belong to a different community, nodes in the network can have multiple community affiliations. Unlike other methods that resolve overlapping communities in which a small fraction of nodes participate weakly in multiple communities \cite{najafi2016overlapping, yeo2014estimates}, edge communities result in ``pervasive overlap,'' such that nodes are tiled by multiple overlapping communities. This notion of pervasive overlap agrees with our understanding of brain function; rather than subtending a narrow or singular set of functions, even traditionally ``unimodal'' brain areas respond to a range of diverse stimuli and conditions \cite{stringer2019spontaneous, allen2019thirst}. In Faskowitz et al \cite{faskowitz2019edge}, entropy measures were used to quantify the level of overlap in each brain region, which revealed heterogeneous patterns. Interestingly, the highest levels of overlap were observed in sensorimotor and attentional networks, suggesting that these regions may play a previously undisclosed role in supporting a wide range of cognitive functions.
	
	In summary, the edge-centric approach provides another lens through which we can study the brain's organization, function, and dynamics. Just as SC and FC offer complementary insight, so does edge connectivity, revealing changes in co-fluctuation patterns over fast timescales and characterizing the brain's higher-order interactions. The edge-centric approach presents multiple opportunities for future studies. First, we can take advantage of the fact that edge time series decompose FC into its framewise contributions to reconstruct FC using only select subsets of frames. These subsets can be chosen in a specific way so as to maximize the relationship of the reconstructed FC with behavior or to enhance group differences. eFC, itself, appears to be useful in amplifying subject-specific features of networks \cite{jo2020subject}, opening up the possibility that this added personalization will improve studies of individual differences.

	\section*{Conclusion}
	
	A perfect confluence of factors -- availability of data, analytic framework, and computational resources -- has created an environment that allows network neuroscience and connectomics to flourish. These new disciplines provide a quantitative framework for exploring patterns of brain connectivity in health and disease and across a vast range of spatiotemporal scales. This framework is beginning to reveal some of the key features of brain networks -- small-world architecture for efficient information processing, modules to support specialized information processing, hubs and rich clubs for integrating information between modules, and a strong dependence on spatial layout that helps reduce the brain's material and metabolic cost of wiring. The future of connectomics includes powerful new methodologies for uncovering  network-level mechanisms and controlling brain activity while new recording and imaging techniques make it possible to resolve networks at increasingly finer spatial scales but with the benefit of broader brain coverage. While the advanced concepts introduced here are necessarily new and continue to be explored and refined within the field of network neuroscience proper, time-varying connectivity, network control, and edge-centric connectomics hold promise in applied fields, such as connectomic deep brain stimulation. It is likely, then, that in future studies, these and other cutting edge methods from network neuroscience will move from exercises in theory and into application.
	
	\bibliography{../biblio_connectomics}

\end{document}